\documentclass[a4paper]{jpconf}
\usepackage{graphicx, color}
\begin{document}
\title{Experimental results on top exotic (non-SUSY) from the LHC}

\author{Florencia Canelli}

\address{Physik Institut, Universit\"at Z\"urich}

\ead{canelli@physik.uzh.ch}

\begin{abstract}
This proceeding describes the experimental results presented at the TOP2014 conference on searches for physics beyond the standard model in final states containing top quarks by the ATLAS and CMS experiments at the LHC. The searches presented were done in the context of a wide range of theoretical models except those related to Supersymmetry.  Results presented use data taken with about 20 fb$^{-1}$ of 8 TeV proton-proton collisions from the LHC.  No significant excesses beyond the standard model are observed, therefore limits on potential new signals are set. 

\end{abstract}

\section{Introduction}

Since the top quark mass is the largest in the standard model, it enters theories as the largest quantum correction. Many models beyond the standard model  hold a special role for top quarks.  Here we present results for searches for physics beyond the standard model in models not involving supersymmetry.  All searches use proton-proton collisions at 8 TeV, most of them use the full data set of LHC Run I (about 20 fb$^{-1}$) collected by the ATLAS and CMS experiments. The talk is divided in three sections according to the type of physics models and analysis techniques: searches for new resonances and heavy bosons,  searches for dark matter produced in association with top quarks, and searches for new vector-like fermions. 

\section{Resonances decaying to top quarks}
An extended gauge sector featuring massive charged bosons, $W'$ and $Z'$, is predicted in many extensions of the standard model \cite{wprime-theory}.
Searches for a $W'$ boson, a heavy partner of the standard model $W$ boson, are traditionally done in the lepton plus neutrino final state. The current limit on this search is 3.74 GeV. This limit assumes the neutrino is lighter than the $W'$. Instead, if the $W'$ boson couples to a right-handed neutrino which is heavier than the $W'$, it would decay to top and bottom quarks.  Searches are also performed for left-handed $W'$ bosons where the interference between the $W'$ and the s-channel single-top production is also taken into account. 

\begin{figure}[h]
\begin{minipage}{18pc}
\begin{center}
\includegraphics[width=12pc]{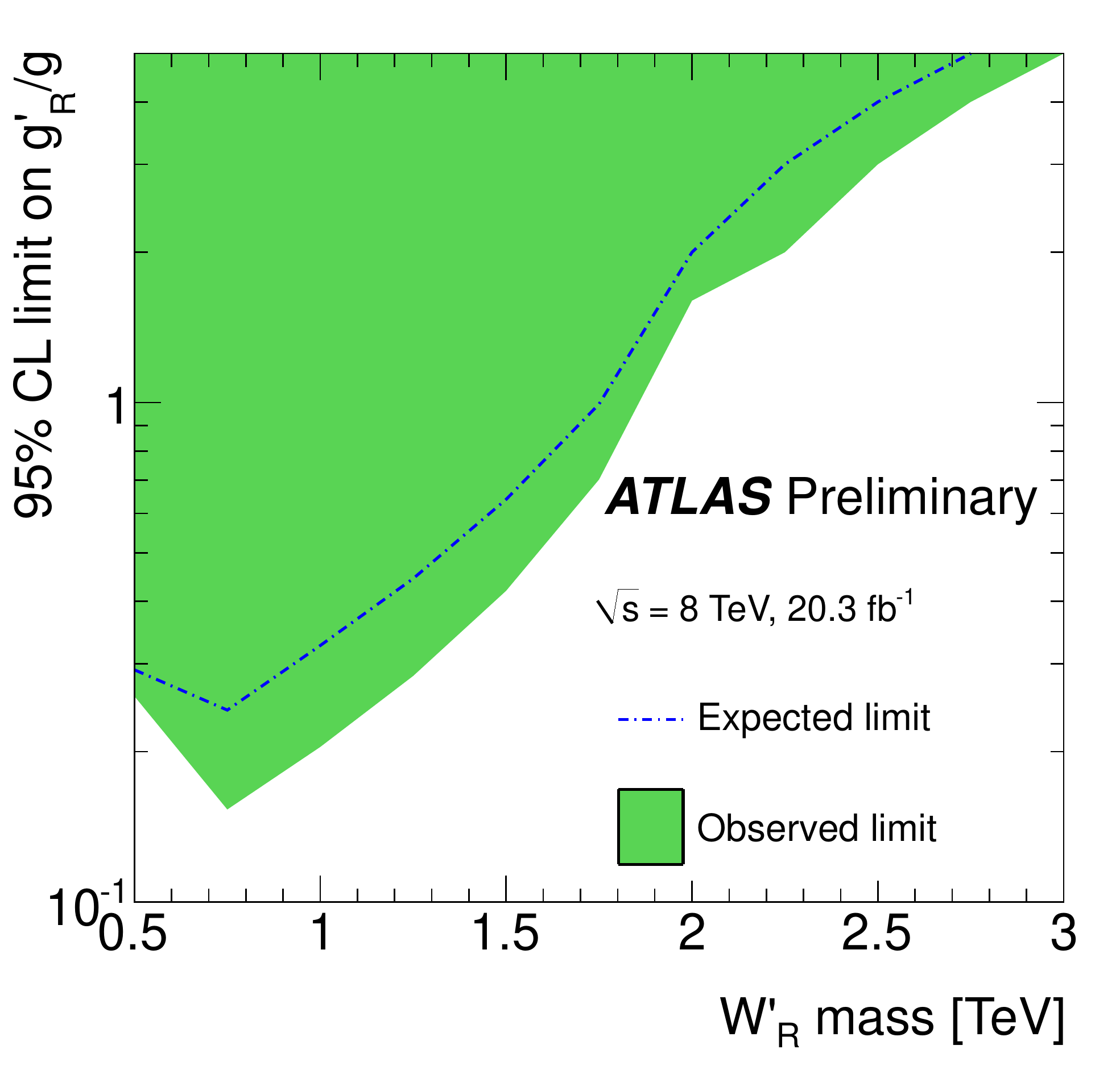}
\caption{\label{wprime-coupling}\small{Limits on the coupling strength for different $W'_R$ boson mass hypotheses (ATLAS).}}
\end{center}
\end{minipage}\hspace{2pc}%
\begin{minipage}{18pc}
\begin{center}
\includegraphics[width=14pc]{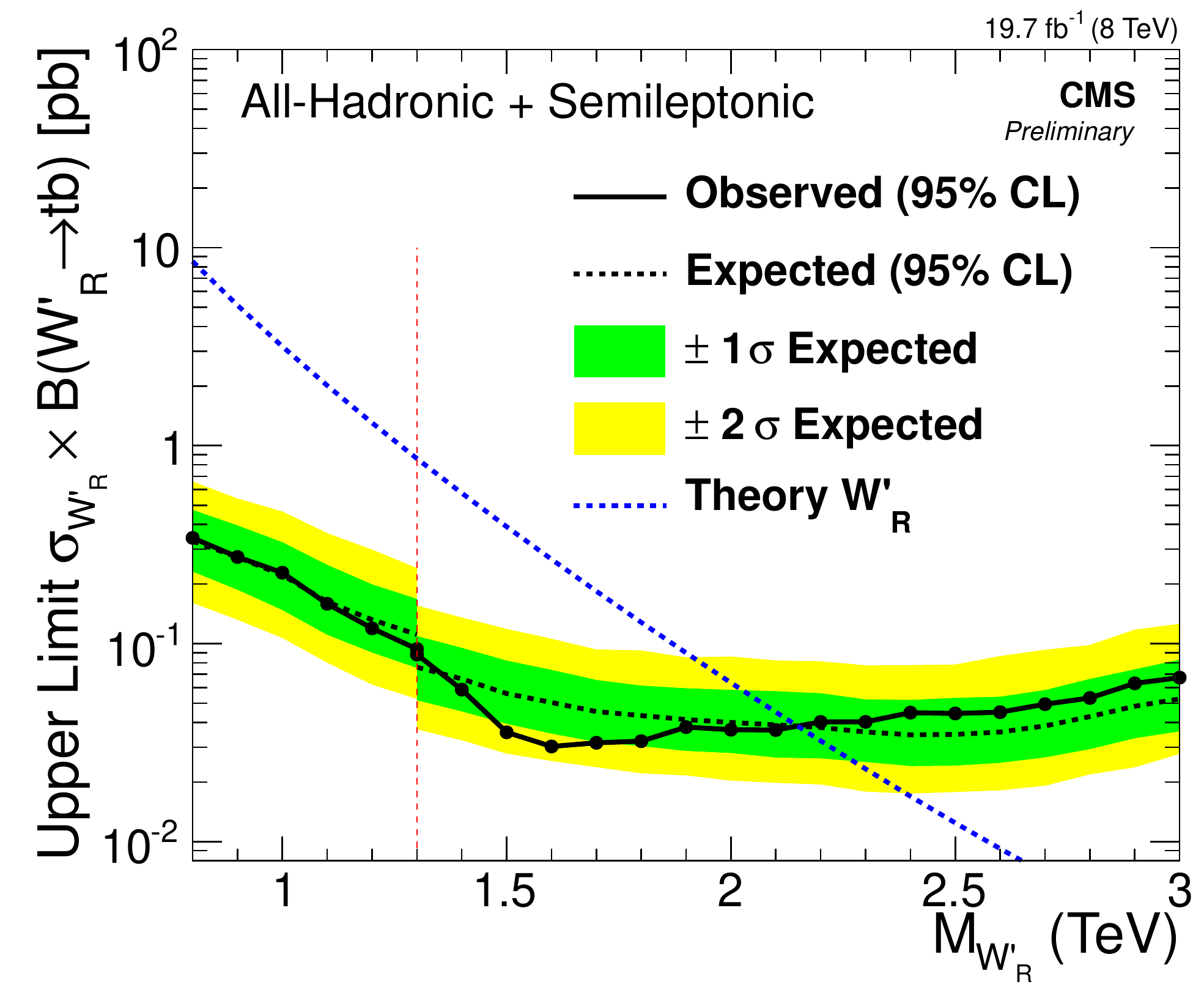}
\caption{\label{wprime-combo}\small{95\% C.L. upper limit on the production of $W'_R$ for different $W'_R$ boson mass hypotheses (CMS).}}
\end{center}
\end{minipage} 
\end{figure}

Table \ref{wprime-ljets} shows the expected and observed limits as a function of $W'$ boson mass for $W' \to tb$ in the final state with a lepton, missing transverse energy,  jets and at least one $b$-tag for the CMS and ATLAS experiments. Results from the CMS experiment were obtained by fitting the invariant mass of the reconstructed top and bottom quarks \cite{CMS-wprime1}.  In this proceeding, ATLAS presents new preliminary results using a multivariate boosted decision tree approach  (Fig.~\ref{wprime-coupling}) \cite{ATLAS-wprime1}.  A more detailed comparison of the CMS and ATLAS analyses started at the TOP2014 conference.  Currently studies indicate that CMS and ATLAS are using a different theoretical assumptions and their limits might not be directly comparable.

\begin{center}
\begin{table}[h]
\caption{\label{wprime-ljets}C.L. 95\% limits on the $W'$ boson mass in TeV for $W' \to tb \to$ lepton+jets  using a dataset of about 20 fb$^{-1}$.}
\centering
\begin{tabular}{@{}*{7}{l}{|}}
\br
Model&CMS Observed (Expected)& ATLAS Observed (Expected)\\
\mr
$W'_L$&1.84 (1.84)& 1.70 (1.54)\\
$W'_R$&2.13 (2.12)& 1.92 (1.75)\\
\br
\end{tabular}
\end{table}
\end{center}

\vspace{-0.3cm}

Searches for the $W'$ boson were also done in the all-jets final state of $W' \to tb$. ATLAS and CMS use techniques for reconstructing highly boosted objects in order to reconstruct the invariant mass of the top and bottom quark system. The multijet background estimate is derived from data.  Table \ref{wprime-alljets} shows the current limits for ATLAS and CMS \cite{ATLAS-wprime2,CMS-wprime2}.

\begin{center}
\begin{table}[h]
\caption{\label{wprime-alljets}C.L. 95\% limits on the $W'$ mass in TeV for $W' \rightarrow tb \rightarrow$ all-jets using a dataset of about 20 fb$^{-1}$.}
\centering
\begin{tabular}{@{}*{7}{l}{|}}
\br
Model&CMS Observed (Expected)& ATLAS Observed (Expected)\\
\mr
W'$_L$&1.94 (NA)& 1.68 (1.63) [no SM interefence considered]\\
W'$_R$&2.02 (1.99)& 1.76 (1.85)\\
\br
\end{tabular}
\end{table}
\end{center}
\vspace{-0.3cm}
Currently, the most stringent limits are produced by the CMS experiment by combining the lepton+jets and all-hadronic channels resulting in a limit of 2.15 TeV for $W'_R$ (Fig.\ref{wprime-combo}) \cite{CMS-wprime2}. 

Many theories beyond the standard model predict the existence of new particles that decay to  $t\bar t$ quarks. Searches for resonant production of $t\bar t$ quarks can test theories such as top-color, chiral color models, Randall-Sundrum (RS) models with warped extradimensions, etc. by looking for a bump in the $t\bar t$ invariant mass spectrum \cite{zprime-theory}. Two benchmark models are tested: 
\begin{itemize}
\item a narrow resonance in comparison with detector resolution (topcolor, leptophobic, $Z’$) with $\Gamma_{Z'}/m_{Z’}$=1.2\% (or 1\%) with a K-factor of 1.3.
\item a broad resonance (Kaluza Klein gluons from RS models with extradimensions, $g_{KK}$) with $\Gamma_{Z'}/m_{Z’}$=15.3\% (or 10-15\%) with no K-factor.
\end{itemize}

The analyses presented use techniques for resolving boosted objects to reconstruct the $t \bar{t}$  pair and fit its invariant mass distribution to find a signal. Results from ATLAS analyzing the $t \bar t$ single lepton decay channel using resolved and boosted techniques are shown in Table \ref{zprime} \cite{ATLAS-zprime}. Note that this result uses a smaller dataset of 14fb$^{-1}$ in comparison to the other results presented in this proceeding. CMS results shown in Table \ref{zprime} are extracted after combining the results from the single-lepton and all-hadronic analyses \cite{CMS-zprime}. 

\begin{center}
\begin{table}[h]
\caption{\label{zprime}C.L. 95\% limits on $Z' \rightarrow t\bar t$.}
\centering
\begin{tabular}{@{}*{7}{l}{|}}
\br
Model&ATLAS Observed (Expected)& CMS Observed (Expected)\\
 &$t\bar t \rightarrow$ lepton+jets ($L$=14 fb$^{-1}$) & Z' $t\bar t \rightarrow$ lepton+jets + all-jets ($L$=20 fb$^{-1}$)\\
\mr
$Z'$ ($\Gamma_{Z'}/m_{Z’}$=1.2\%) & 1.8 (1.9)& 2.1 (2.1)\\
$g_{KK}$ ($\Gamma_{Z'}/m_{Z’}$=15.3\%)& 2.0 (2.1)& 2.5 (2.4)\\
\br
\end{tabular}
\end{table}
\end{center}
\vspace{-0.3cm}

\section{Dark matter production in association with top quarks}

\begin{figure}[h]
\begin{minipage}{12pc}
\includegraphics[width=12pc]{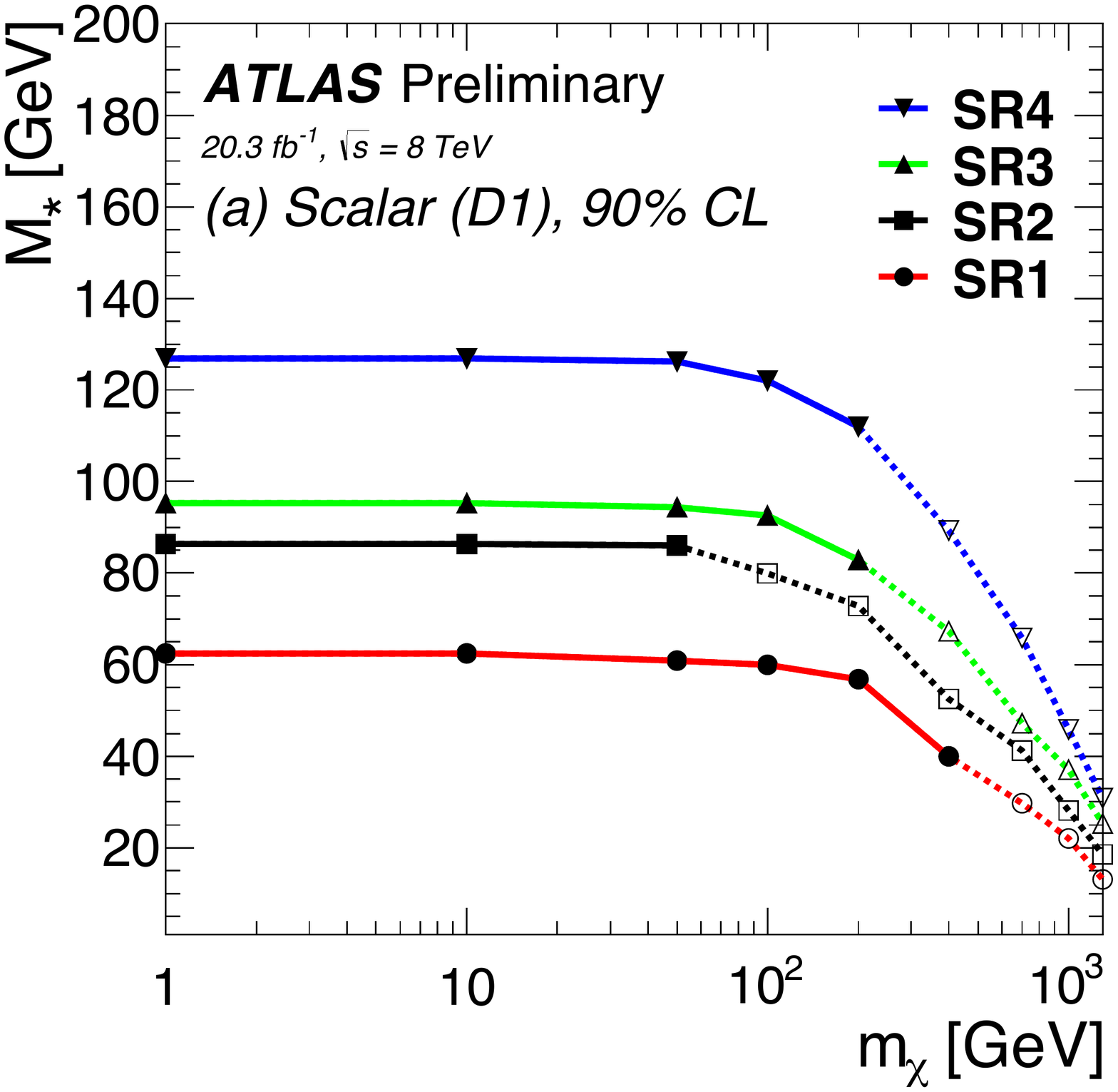}
\end{minipage}
\begin{minipage}{12pc}
\includegraphics[width=12pc]{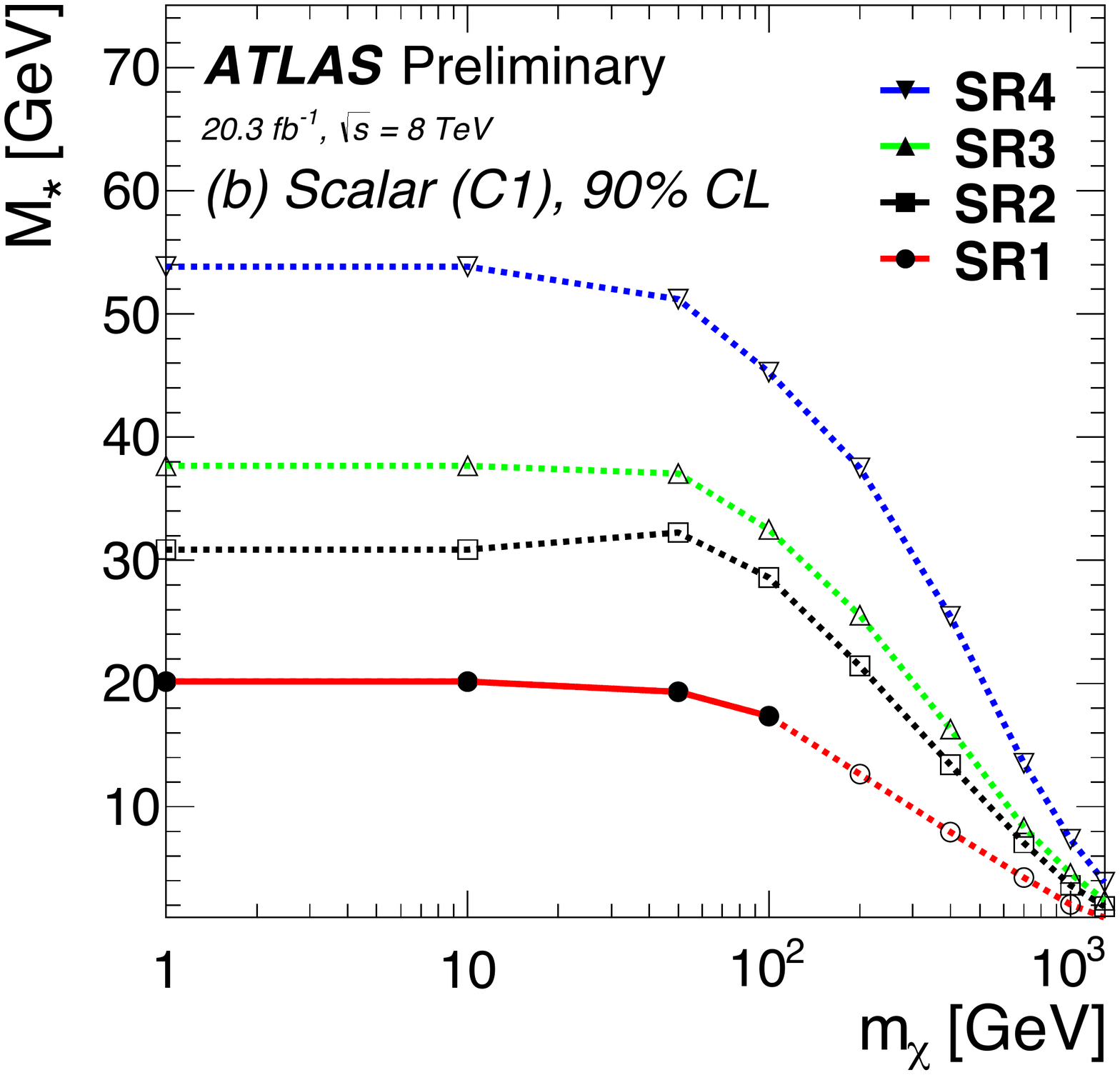}
\end{minipage} 
\begin{minipage}{12pc}
\includegraphics[width=12pc]{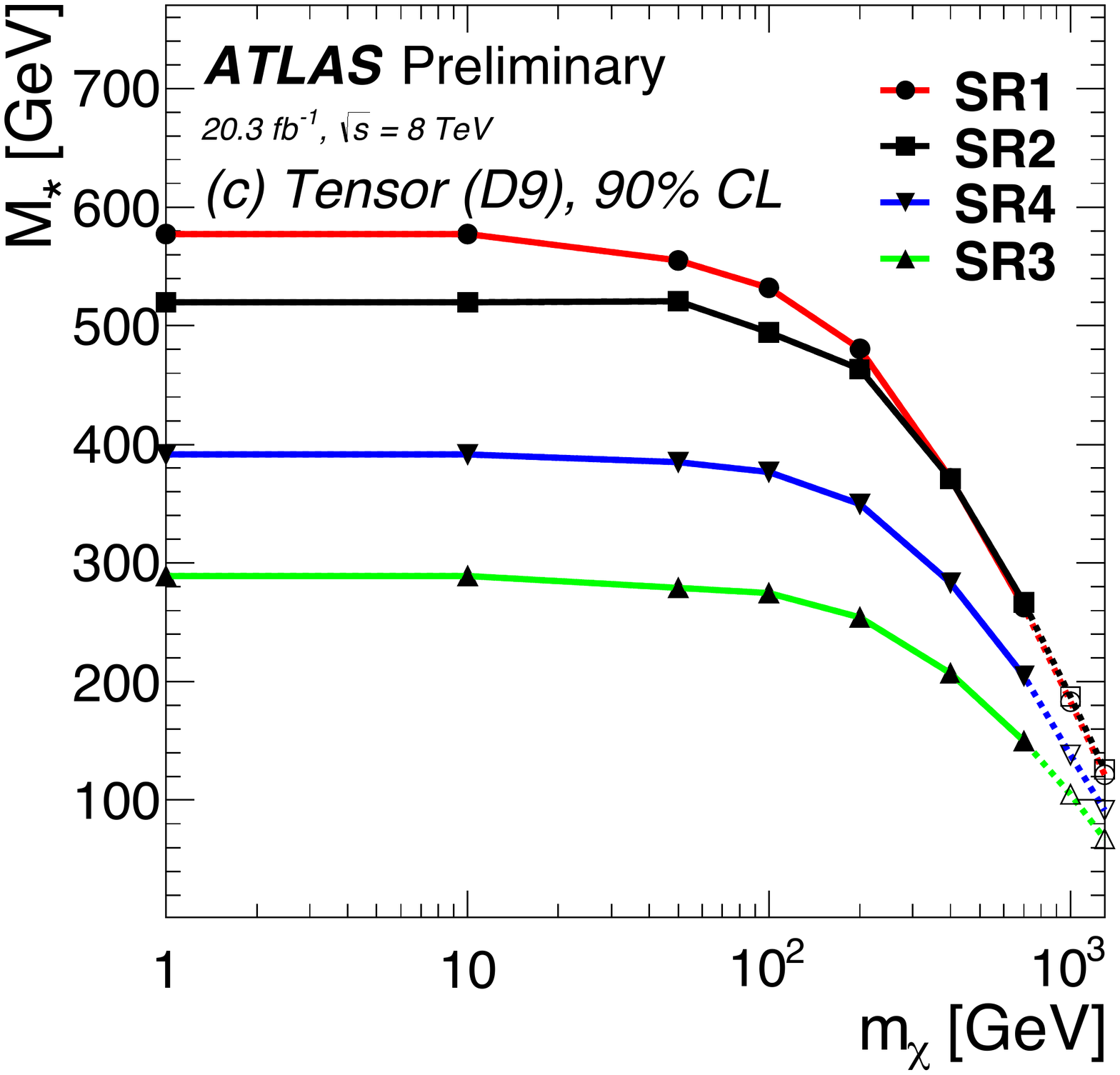}
\end{minipage} 
\caption{\label{dm-atlas}\small{95\% C.L. upper limit on the effective mass scale $M^{*}$ of different operators for different signal regions defined as SR1 and SR2 for  1 or 2 b-tags and no lepton, SR3 for $t\bar t$ with all-hadronic decays with more than 5 jets, and SR4 for $t\bar t$ with a single lepton decay and $b$-tags. Solid curves correspond to $m_{\chi}<Q_{TR}$ (momentum transferred).} }
\end{figure}

Searches for dark matter (DM) particles produced in a collider are most sensitive to low masses ($m_{\chi}<$10 GeV)  and provide complementary information to to direct detection searches, which are most sensitve to larger DM masses.  If particles that mediate the interaction between dark matter and standard model particles ($M^{*}$) are too heavy to be produced, interactions can be described as contact operators in an effective field theory (EFT) with operators expressed in terms of $M^{*}$.   Different operators are tested using models where the dark matter is produced with a top quark pair \cite{dm-theory}. The CMS experiment sets lower limits on the interaction of scale $M^{*}$ assuming a D1 coupling in the single-lepton and dilepton channel \cite{CMS-dm1,CMS-dm2}. ATLAS presents a new result using single-lepton and all-jets $t\bar t$ decays and sets lower limits on three different operators, D1, D11, D9 \cite{ATLAS-dm1}. These new results are shown in Fig. \ref{dm-atlas}. 

\begin{figure}[h]
\begin{minipage}{14pc}
\begin{center}
\includegraphics[width=14pc]{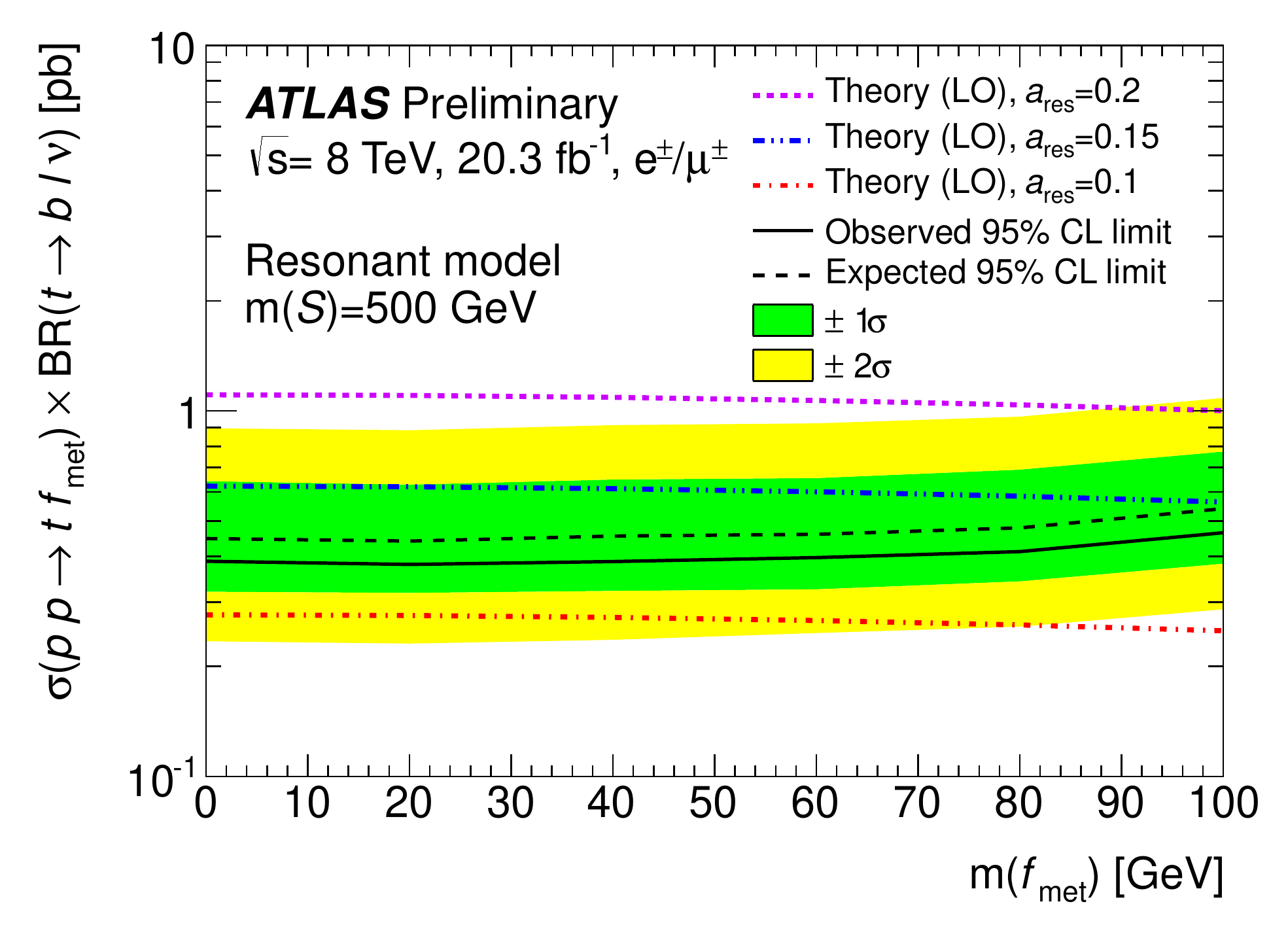}
\end{center}
\end{minipage}
\begin{minipage}{14pc}
\begin{center}
\includegraphics[width=14pc]{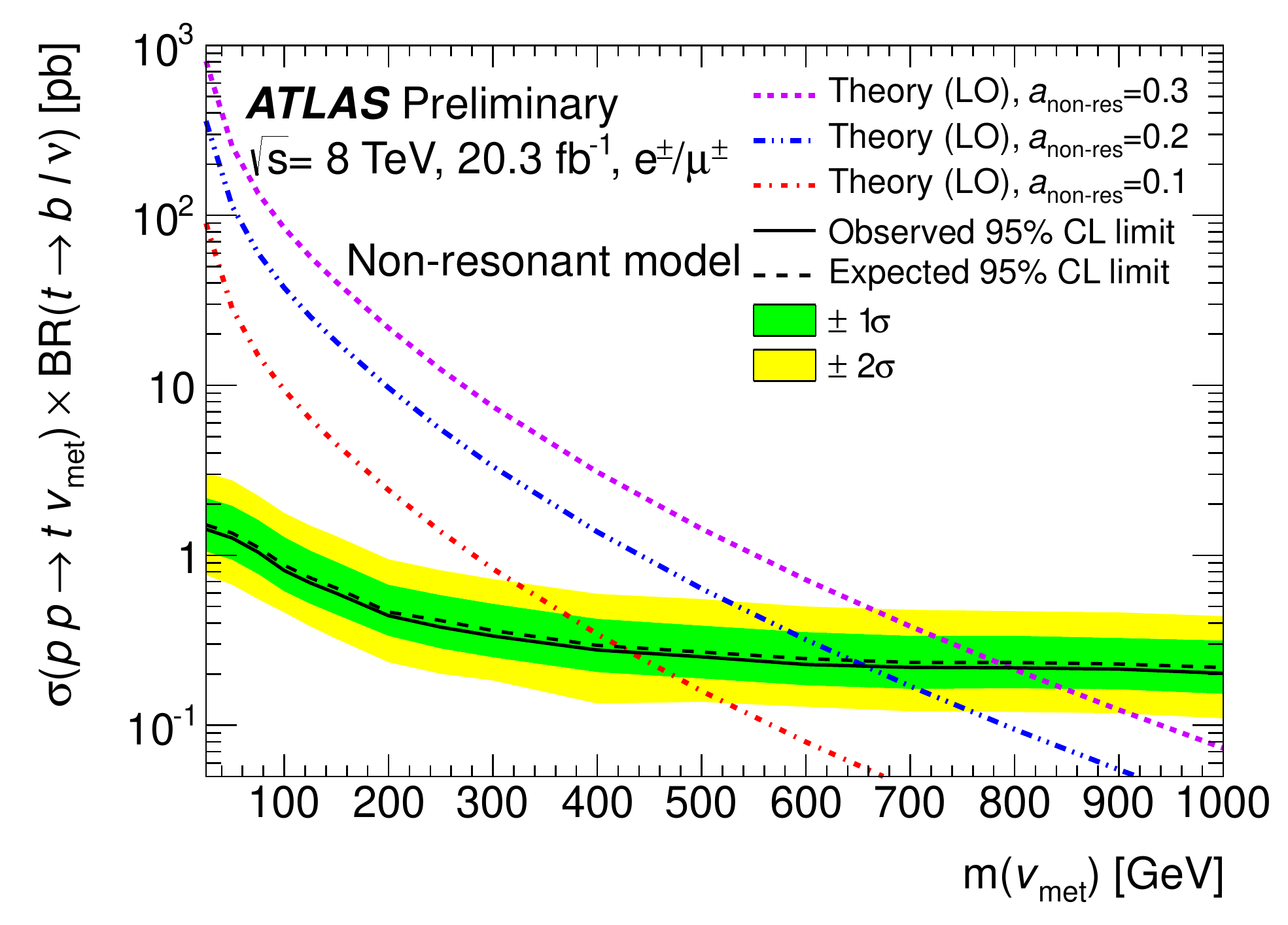}
\end{center}
\end{minipage} 
\caption{\label{monotop-atlas}\small{Expected and observed excluded cross-section times branching ratio limits for the resonant (left) and non-resonant (right) models, as a function of the mass of $f_{met}$ and $v_{met}$, respectively. The predicted signal cross-sections for different coupling values are also shown.}}
\end{figure}

Searches for missing energy produced with a single top quark (mono-top) have been done by CMS and ATLAS. CMS extracts limits using the reconstructed top quark invariant mass in final states with three jets and large missing transverse energy and set limits within the EFT interpretation where the dark matter particles are scalar and vector bosons, excluding scalar (vector) DM particle masses below 327 (655) GeV \cite{CMS-dm3}.  Newly presented at this conference were the limits on mono-top searches by the ATLAS experiment using the single-lepton channel \cite{ATLAS-dm2}. The search was interpreted with two models: a $resonant$ production of a +2/3 charged spin-0 boson, S, decaying into a right-handed top quark and a non-interacting neutral spin-$\frac{1}{2}$ fermion, $f_{met}$, and  as a $non-resonant$ production of a non-interacting neutral spin-1 boson, in association with a right-handed top quark, $v_{met}$. Results are shown in Fig. \ref{monotop-atlas}. 

\section{Vector-like quarks}

Vector-like quarks are defined as quarks with left and right handed components transforming identically under (SU2)$_L$.  Vector-like quarks (VLQ) have been proposed in many theories beyond the standard model, mainly to address the naturalness problem in models like little Higgs, composite Higgs,  etc.  They may exist as weak-isospin doublets, singlets, or triplets and can occur as down-type quark ($B$), up-type quark ($T$), or as a quark with even more exotic electric charges of 4/3 e or 5/3 e.  Predominantly they decay to third generation quarks, and therefore produce a variety of search final states. Searches are performed for pair or single production of $B\rightarrow Wt, Zb, Hb$ and $T \rightarrow Wb, Zt, Ht$. Pair production searches are more common as the production of single $B$ and $T$ quarks via the weak interaction has lower cross sections.  In general signatures have large number of jets, leptons, and b-jets. Some of the final states have leptons with the same-sign charges. Selection of same-sign leptons is exploited to suppress SM backgrounds.  Top quarks or bosons decaying from heavy quarks can have large momenta, showing a boosted signature. Many CMS and ATLAS analyses use boosted tagging techniques to identify Higgs, top quark, $W$ or $Z$ bosons from heavy quarks. 

The search strategy consists in discriminating signal from background using  $H_T$, the scalar sum of the transverse momenta of the reconstructed jets, leptons, and missing transverse energy, or $M_{reco}$, the reconstructed invariant mass of the final-state objects in each channel.  No new results were presented at this conference. In absence of an observed signal the experiments produce limits on different couplings and VLQ masses.  Depending of the decay mode, $B$ and $T$ quarks are ruled out between 600 to 800 GeV. ATLAS and CMS results can be found in references \cite{ATLAS-vlq,CMS-vlq1}. In addition searches for quarks with a charge of 5/3 e and 2/3 e have not seen any significant deviation from the SM and limits exclude top quark partners with masses below 800 GeV for all possible values of the branching fractions \cite{CMS-vlq2}.



\section{Conclusions}

The top quark plays an important role in beyond-standard model (BSM) theories. ATLAS and CMS have used 8 TeV proton-proton collisions from the LHC and the full dataset of about 20 fb$^{-1}$ to search for different BSM signatures. Here we have presented searches for $W'$ and $Z'$ bosons as well as $g_{KK}$, production of dark matter in association with single top quark and top quark pairs, and lastly different searches for vector-like quarks produced singly or in pair. No hint for new physics was found but stringent limits on these models have been set.  All the top quark decay channels were explored and many different reconstruction and analysis techniques were developed and improved for these searches.


\section*{References}
\begin{thebibliography}{9}
\bibitem{wprime-theory} G. Burdman, B. A. Dobrescu, and E. Ponton, Phys. Rev. D74 (2006) 075008, arXiv:hep-ph/0601186; H.-C. Cheng, C. T. Hill, S. Pokorski, and J. Wang, Phys. Rev. D64 (2001) 065007, arXiv:hep-th/0104179; T. Appelquist, H.-C. Cheng, and B. A. Dobrescu, Phys. Rev. D64 (2001) 035002, arXiv:hep-ph/0012100; J. C. Pati and A. Salam, Phys. Rev. D10 (1974) 275–289; R. N. Mohapatra and J. C. Pati, Phys. Rev. D11 (1975) 566–571; G. Senjanovic and R. N. Mohapatra, Phys. Rev. D12 (1975) 1502; M. Perelstein, Prog. Part. Nucl. Phys. 58 (2007) 247–291, arXiv:hep-ph/0512128; D. J. Muller and S. Nandi, Phys. Lett. B383 (1996) 345–350, arXiv:hep-ph/9602390; E. Malkawi, T. M. Tait, and C. Yuan, Phys. Lett. B385 (1996) 304–310, arXiv:hep-ph/9603349.

\bibitem{CMS-wprime1} CMS collaboration, JHEP 05 (2014) 108.

\bibitem{ATLAS-wprime1} ATLAS collaboration, submitted to Phys. Lett. B, arXiv:1410.4103.

\bibitem{ATLAS-wprime2} ATLAS collaboration, submitted to Eur. Phys. J. C, arXiv:1408.0886.

\bibitem{CMS-wprime2} CMS collaboration, CMS-PAS-B2G-­12-­009.

\bibitem{zprime-theory} N. Arkani-Hamed, S. Dimopoulos, and G. R. Dvali, Phys. Lett. B 429, 263 (1998); L. Randall and R. Sundrum, Phys. Rev. Lett. 83, 3370
(1999); L. Randall and R. Sundrum, Phys. Rev. Lett. 83, 4690 (1999); K. Agashe, A. Belyaev, T. Krupovnickas, G. Perez, and J. Virzi, Phys. Rev. D 77, 015003 (2008); N. Arkani-Hamed, A. G. Cohen, and H. Georgi, Phys. Lett. B 513, 232 (2001); J. L. Rosner, Phys. Lett. B 387, 113 (1996); R. M. Harris and S. Jain, Eur. Phys. J. C 72, 2072 (2012); P. H. Frampton and S. L. Glashow, Phys. Lett. B 190, 157 (1987); D. Choudhury, R. M. Godbole, R. K. Singh, and K. Wagh, Phys. Lett. B 657, 69 (2007); D. Dicus, A. Stange, and S. Willenbrock, Phys. Lett. B 333, 126 (1994); R. Frederix and F. Maltoni, J. High Energy Phys. 01 (2009) 047.

\bibitem{ATLAS-zprime} ATLAS collaboration,  ATLAS-CONF-2013-052.

\bibitem{CMS-zprime} CMS collaboration, Phys. Rev. Lett. 111, 211804.

\bibitem{dm-theory} K. Cheung et al., JHEP 2010 (2010)  081; M. Beltran et al., JHEP 2010 (2010) 1–17; J. Goodman et al., Phys. Rev. D 82 (2010) 116010; P. J. Fox, R. Harnik, J. Kopp, and Y. Tsai, Phys. Rev. D 85 (2012) 056011; P. J. Fox, R. Harnik, R. Primulando, and C.-T. Yu, Phys. Rev. D 86 (2012) 015010; A. Rajaraman, W. Shepherd, T. M. P. Tait, and A. M. Wijangco, Phys. Rev. D 84 (2011) 095013.

\bibitem{CMS-dm1} CMS collaboration, CMS-PAS-B2G-13-004,

\bibitem{CMS-dm2} CMS collaboration, CMS-PAS-B2G-14-004.
\bibitem{CMS-dm3} CMS collaboration, submitted to Phys. Rev. Lett., arXiv:1410.1149.
\bibitem{ATLAS-dm1} ATLAS collaboration, submitted to Eur. Phys. J. C, arXiv:1410.4031.
\bibitem{ATLAS-dm2} ATLAS collaboration, submitted to Eur. Phys. J. C, arXiv:1410.5404.

\bibitem{ATLAS-vlq} ATLAS collaboration, JHEP 11 (2014) 104, ATLAS-­CONF-­2013-­060, ATLAS-CONF-2013-056, ATLAS-­CONF-­2013-­051, ATLAS-­CONF-­2013-­018.

\bibitem{CMS-vlq1} CMS collaboration, CMS-PAS-B2G-14-001, CMS-PAS-B2G-12-020, CMS-PAS-B2G-14-003, CMS-PAS-B2G-14-002, CMS-PAS-B2G-12-008, CMS-PAS-B2G-13-003, CMS-PAS-B2G-12-019, CMS-PAS-B2G-12-021. 

\bibitem{CMS-vlq2} CMS collaboration, Phys. Lett. B 729 (2014) 149,  Phys. Rev. Lett. 112 (2014) 171801.

\end{thebibliography}
\end{document}